\pdfoutput=1

\documentclass[11pt]{article}

\usepackage{ACL2023}
\usepackage{tikz}   
\usepackage[most]{tcolorbox}
\usepackage{times}
\usepackage{latexsym}
\usepackage{multirow}
\usepackage{pifont}
\usepackage{xcolor} 
\usepackage{soul}
\sethlcolor{lightgreen}
\definecolor{lightpurple}{RGB}{230, 221, 246}
\definecolor{lightgreen}{RGB}{224, 242, 213}
\definecolor{lightgray}{RGB}{211, 211, 211}
\definecolor{lightred}{RGB}{255, 182, 193}
\definecolor{lightblue}{RGB}{212, 235, 255}
\newcommand{\colorlightgreen}[1]{\sethlcolor{lightgreen}\hl{#1}}

\newcommand{\colorlightblue}[1]{\sethlcolor{lightblue}\hl{#1}}
\usepackage{colortbl}
\usepackage{tcolorbox}
\usepackage{bbding}
\usepackage{amsmath}
\usepackage{graphicx}
\usepackage{courier}  
\usepackage{makecell}
\usepackage{appendix}
\usepackage{natbib}  
\usepackage{caption} 
\usepackage[T1]{fontenc}

\usepackage[utf8]{inputenc}

\usepackage{microtype}
\usepackage{hyperref}
\usepackage{inconsolata}
\newtcolorbox{prompt}[1]{colback=gray!5,colframe=gray!35!black,fonttitle=\bfseries, title={#1}}
\definecolor{antiquebrass}{rgb}{0.8, 0.58, 0.46}

\title{CMIE: Combining MLLM Insights with External Evidence for Explainable Out-of-Context Misinformation Detection}

\author{Fanxiao Li\textsuperscript{1,3}, Jiaying Wu\textsuperscript{2}, Canyuan He\textsuperscript{3}, Wei Zhou\textsuperscript{3} \thanks{\quad The corresponding author.}\\
  \textsuperscript{1}School of Information Science and Engineering, Yunnan University\\
   \textsuperscript{2} National University of Singapore \\
   \textsuperscript{3}Engineering Research Center of Cyberspace, Yunnan University \\
    \texttt{lifanxiao15@outlook.com, jiayingwu@u.nus.edu, zwei@ynu.edu.cn} 
  }

\begin{document}
\maketitle
\begin{abstract}
Multimodal large language models (MLLMs) have demonstrated impressive capabilities in visual reasoning and text generation. While previous studies have explored the application of MLLM for detecting out-of-context (OOC) misinformation, our empirical analysis reveals two persisting challenges of this paradigm. Evaluating the representative GPT-4o model on direct reasoning and evidence augmented reasoning, results indicate that MLLM struggle to capture the deeper relationships—specifically, cases in which the image and text are not directly connected but are associated through underlying semantic links. Moreover, noise in the evidence further impairs detection accuracy.
To address these challenges, we propose CMIE, a novel OOC misinformation detection framework that incorporates a Coexistence Relationship Generation (CRG) strategy and an Association Scoring (AS) mechanism. CMIE identifies the underlying coexistence relationships between images and text, and selectively utilizes relevant evidence to enhance misinformation detection. Experimental results demonstrate that our approach outperforms existing methods. Data and code are available at: \url{https://github.com/fanxiao15/CMIE}.
\end{abstract}

\section{Introduction}
The proliferation of multimodal misinformation on social media poses significant risks, including impairing public judgment, inciting panic, and causing economic losses \cite{DBLP:journals/corr/harmful_1, harmful_2}.
Multimodal misinformation manifests in various forms, such as image-text mismatches \cite{DBLP:conf/mm/MM-cross, DBLP:conf/emnlp/SEN, CCN, DBLP:journals/ijmir/VERITE} and manipulated media \cite{DBLP:conf/aaai/video_1, DBLP:journals/corr/video_2}. Among image-text mismatches, out-of-context (OOC) misinformation is particularly prevalent \cite{DBLP:conf/emnlp/newsclippings}, where an unaltered image is paired with a new but misleading textual context. Detecting such misinformation is challenging, as repurposing images with different textual descriptions can create seemingly strong yet deceptive correlations.

\begin{figure}[t!]
\begin{center}
    \includegraphics[width=\linewidth]  {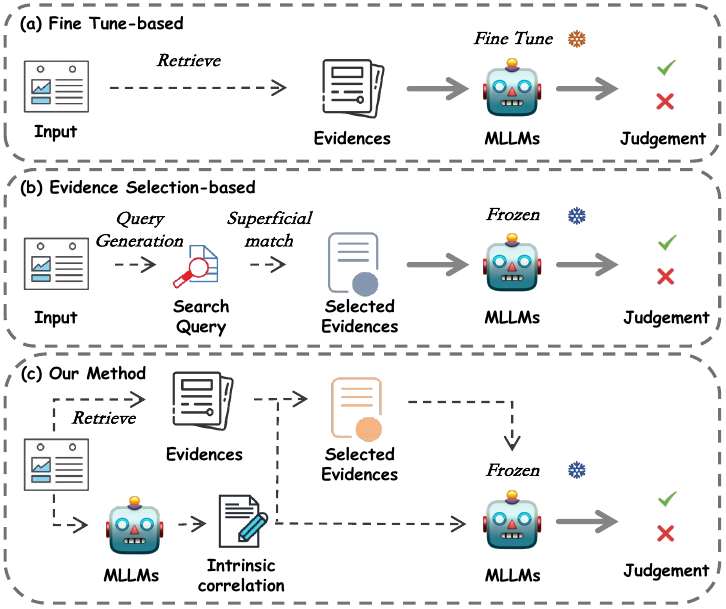}
    \caption{Comparison of different paradigms for MLLM-based OOC misinformation detection.}
    \label{different_paradigm}
\end{center}
\end{figure}

Previous studies \cite{DBLP:conf/emnlp/newsclippings, DBLP:conf/mir/DT-transformer, cosmos} typically detect OOC misinformation by training a small language model (SLM) to assess the consistency between image and text representations. Recently, some works \cite{CCN, DBLP:conf/mir/Muller-BudackTD20, DBLP:conf/emnlp/SEN} attempt to verify information authenticity by retrieving external evidence from the Internet. However, these methods often lack explainability, providing only a final classification label without generating clear, human-readable justifications.

The advent of multimodal large language models (MLLMs) \cite{DBLP:journals/corr/gpt-4o, DBLP:journals/corr/Gemini} has driven significant advancements across various vision tasks \cite{DBLP:journals/ftcgv/MLLM_survey1, DBLP:journals/corr/MLLM_survey2}. Their advanced visual reasoning and text generation capabilities enable high-quality, human-readable explanations for OOC misinformation detection. Additionally, their extensive world knowledge supports zero-shot transfer across diverse datasets, enhancing adaptability to novel misinformation instances.

Recent studies \cite{sniffer, LEMMA, MUSE} integrate Retrieval-Augmented Generation (RAG) techniques to further improve misinformation detection by incorporating external evidence. SNIFFER \cite{sniffer} fine-tunes InstructBLIP \cite{DBLP:conf/nips/InstructBLIP} and uses ChatGPT \cite{DBLP:journals/corr/GPT-4} for explanation label generation, but as shown in Figure \ref{different_paradigm} (a), it is computationally expensive and requires extensive labeled data. In contrast, LEMMA \cite{LEMMA} and MUSE \cite{MUSE} avoid fine-tuning and consider evidence relevance during retrieval (Figure \ref{different_paradigm} (b)). However, they primarily rely on superficial lexical matching and fail to explain how retrieved evidence influences MLLM’s reasoning in detecting OOC misinformation.

To thoroughly investigate the capabilities of MLLM in OOC misinformation detection, this paper explores the following research questions:
\textbf{(1) How well can MLLMs perform in misinformation detection without fine-tuning or external evidence? (2) How does external evidence influence MLLM-based misinformation detection?}

Through our empirical analysis using the representative GPT-4o model (detailed in Section \ref{sec:prelim}), we find that MLLMs can achieve competitive performance in misinformation detection relying solely on their intrinsic knowledge, without requiring fine-tuning or external evidence. However, their judgments tend to be conservative, particularly when provided with image-text pairs that share similar contextual backgrounds but lack obvious inconsistencies. When external evidence is introduced, MLLMs often rely on superficial correlations rather than deeper semantic relationships, limiting their ability to make well-informed decisions. Additionally, noisy or irrelevant evidence further constrains performance improvements.

Motivated by these observations, we propose CMIE, an OOC misinformation detection framework designed to uncover deeper connections between image-text pairs and refine evidence selection accordingly, as illustrated in Figure \ref{different_paradigm}(c).  CMIE introduces a \textbf{coexistence relationship generation} (CRG) strategy that leverages MLLMs to identify and extract the underlying relationships between images and text. Since images are expected to support the credibility of the associated text, they should share common expressive elements. These extracted relationships serve as a core validation thread, enabling a more structured approach to misinformation detection. To incorporate high-quality external evidence, CMIE further incorporates an \textbf{association scoring} (AS) mechanism that evaluates external evidence based on its deeper semantic alignment with the identified image-text relationships. Rather than relying on superficial lexical matches, this mechanism ensures that only the most relevant evidence is considered, mitigating the impact of noisy or misleading information.

The key contributions of this paper can be summarized as follows:

\begin{itemize}
\item  \textbf{In-Depth Exploration}: We examined the performance of MLLM in detecting OOC misinformation without fine-tuning and analyzed the impact of integrating external evidence directly into MLLM. Our findings revealed the importance of incorporating external evidence into the task reasonably.

\item  \textbf{Novel Solution}: We introduced CMIE, a novel framework consist of coexistence relationship generation strategy and the association scoring mechanism. This innovation effectively models the relationship between evidence and the task, significantly enhancing MLLM's performance without fine-tuning.
    
\item  \textbf{Effectiveness Verification}: Extensive experiments demonstrate that our method not only generates human-readable explanations but also achieves superior performance compared to state-of-the-art approaches across multiple evaluation metrics.

\end{itemize}

\section{Exploratory Study}
\label{sec:prelim}
\subsection{Task Definition}
This paper seeks to utilize MLLM for detecting OOC misinformation, producing both final predictions and human-readable explanations.
Given an image-text pair $(I, T)$, knowledge $G$, we expect MLLM to produce a label $Y_{p} \subseteq \left\{Real, Fake \right\}$ and a human-readable explanation $Y_{e}$. Formally, this is expressed as: 
\begin{equation}
    Y_{p}, Y_{e} = MLLM((I, T), G),
\end{equation}
which $G$ may include only internal knowledge solely or a combination of internal knowledge and retrieved external evidence.

\subsection{Evaluation Setting}
we evaluate the performance in detecting OOC misinformation of representative MLLM, specifically GPT-4o \cite{DBLP:journals/corr/gpt-4o}.

\textbf{Dataset.} We conduct our experiments on NewsCLIPpings \cite{DBLP:conf/emnlp/newsclippings}, the largest benchmark for OOC misinformation detection, where both images and texts originate from real-world sources. Following prior research, we select the test set from the Merged/Balance subset, which consists of 71,072 samples in the training set, 7,024 in the validation set, and 7,264 in the test set.

\textbf{Metrics.} We measure performance using precision, recall, F1 score, and test accuracy for both real and fake samples.

\subsection{Evaluation Types}
We primarily explored the MLLM's ability without fine-tuning for OOC detection by leveraging its internal knowledge and combining it with external evidence.

\begin{itemize}
\item  \textbf{Direct reasoning with internal knowledge (DR)}: 
In this method, MLLM relies solely on its internal knowledge to detect OOC misinformation. The image-text pair is input directly into the MLLM, which then generates a label (real or fake) and a corresponding explanation.

\item  \textbf{Augmented reasoning with external evidence (AR)}: 
This approach enhances MLLM's reasoning by incorporating external evidence retrieved from the internet. We provide the titles of posts containing the image-text pair, allowing the MLLM to make more informed judgments.
\end{itemize}

\subsection{Findings}
In Table. \ref{exploration}, we report the exploratory results for the two evaluation types mentioned above. Our findings are as follows:

\begin{table}[ht]
\renewcommand\arraystretch{1.2}
\setlength{\tabcolsep}{2 pt} 
\small
  \begin{center}
    \caption{OOC misinformation detection performance of direct reasoning (DR) using internal knowledge and augmented reasoning (AR) with external knowledge.}
\begin{tabular}{cccccc}
\hline \hline
 - & \textbf{Type} & \textbf{Precision} & \textbf{Recall} & \textbf{F1} & \textbf{Accuracy}\\ \hline
 \multirow{2}{*}{\textbf{real}} & \multirow{1}{*}{DR} & 0.79 &  0.89 & 0.84 & 0.83\\ \cline{3-6}
 & \multirow{1}{*}{AR} & 0.86 &  0.83 & 0.84 & 0.85\\ \hline
 \multirow{2}{*}{\textbf{fake}} & \multirow{1}{*}{DR} & 0.87 &  0.76 & 0.81 & 0.83\\ \cline{3-6}
 & \multirow{1}{*}{AR} & 0.83 &  0.86 & 0.85 & 0.85\\ 
\hline \hline

\end{tabular}
    \label{exploration}
  \end{center}
\end{table}

(1) Both DR and AR achieve competitive performance, highlighting \textbf{the feasibility of directly using MLLMs for OOC misinformation detection without fine-tuning.}

\begin{figure}[ht]
\begin{center}
    \includegraphics[width=\linewidth]  {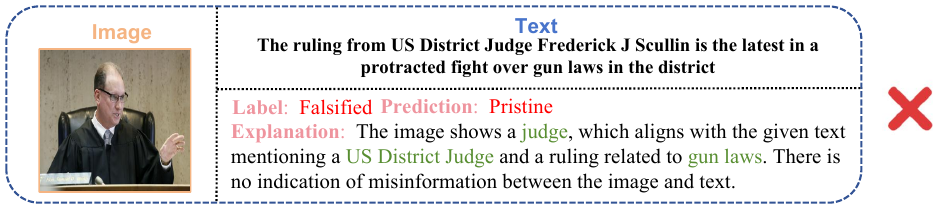}
    \caption{Example of direct reasoning in OOC misinformation detection using MLLM's internal knowledge.}
    \label{direct_intent}
\end{center}
\end{figure}

(2) When relying solely on internal knowledge for detection, \textbf{MLLM tend to produce conservative responses.} For instance, as illustrated in Figure  \ref{direct_intent}, as long as there is a slight connection of the image-text pair and the MLLM cannot identify obvious errors via its internal knowledge, the pair is often considered as real. This observation is supported by the results in Table. \ref{exploration}, which shown a high recall rate for real samples. In OOC misinformation, images are used to enhance the credibility of the texts, leading to a certain degree of correlation of the pair. This correlation complicates the accurate identification of misinformation.

\begin{figure}[ht]
\begin{center}
    \includegraphics[width=\linewidth]  {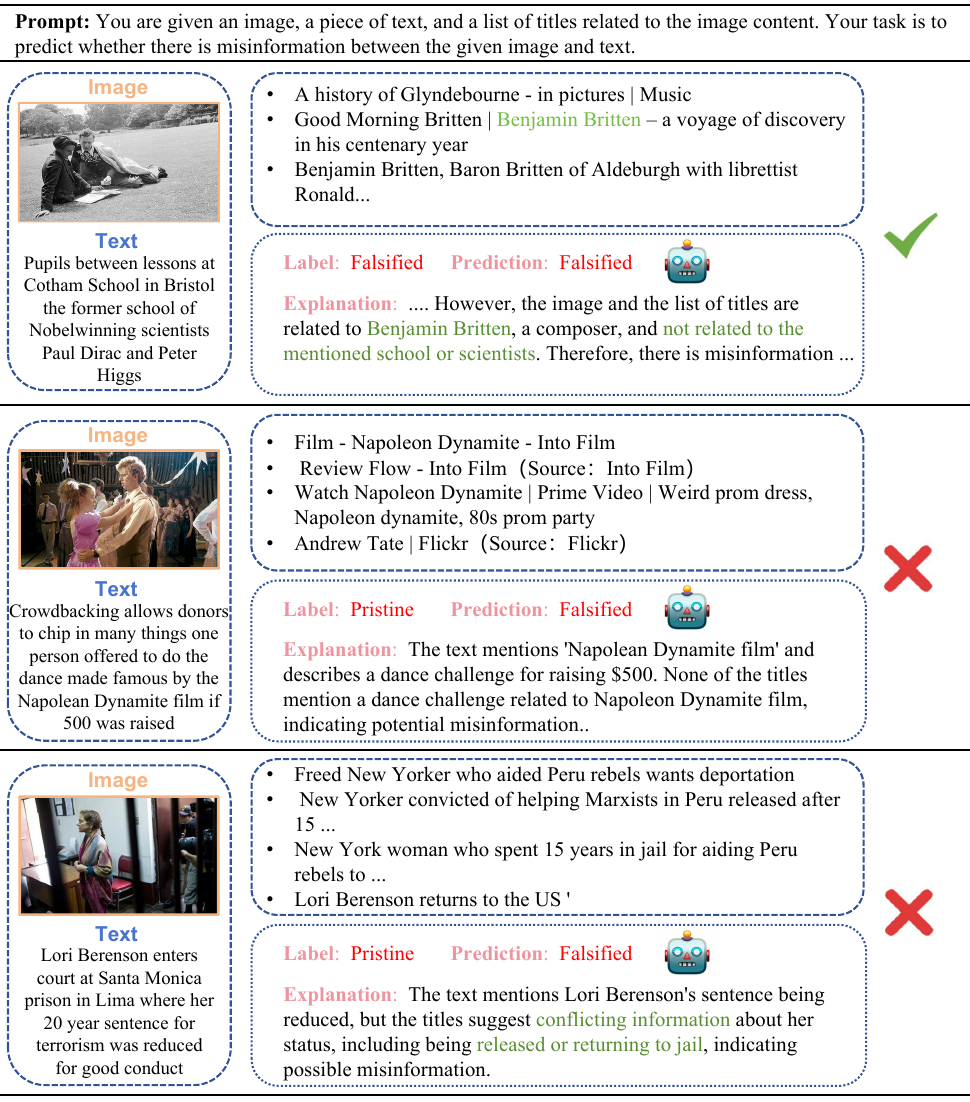}
    \caption{Examples of augmented reasoning by integrating external evidence.}
    \label{RAG_noise}
\end{center}
\end{figure}

 (3) When combining external evidence for augmentation, \textbf{overall performance improves, but the improvement is limited.} The upper part of Figure  \ref{RAG_noise} provides an example where external evidence effectively counteracts incorrect matches. As shown in Table. \ref{exploration}, adding evidence significantly improves the recall rate for fake samples.
However, the improvement is constrained, which we attribute to two main factors. First, MLLM often struggled to grasp the core thread of verification after integrate evidence, as illustrated in the middle part of Figure  \ref{RAG_noise}. When there is no clear correlation between the evidence and the content to be verified, MLLM fails to find implicit relationships of the image-text pair as seen in direct reasoning. Instead, it only determines whether there is a correlation between the image-text pair and the evidence. Second, evidence can be noisy. As shown in the lower part of Figure \ref{RAG_noise}, part of the evidences may be related to the image, but not to the core thread being verified. For effective verification, it is crucial to prioritize highly relevant evidence and treat less relevant evidence as noise.
According these analyses, directly using evidence for reasoning leads to limited performance improvement. Results in Table. \ref{exploration} further illustrated that while incorporating evidence improves average accuracy, other metrics for both real and fake samples show unstable improvements, and may decline due to these aforementioned issues.

\begin{figure*}[ht]
\begin{center}
    \includegraphics[width=\linewidth]  {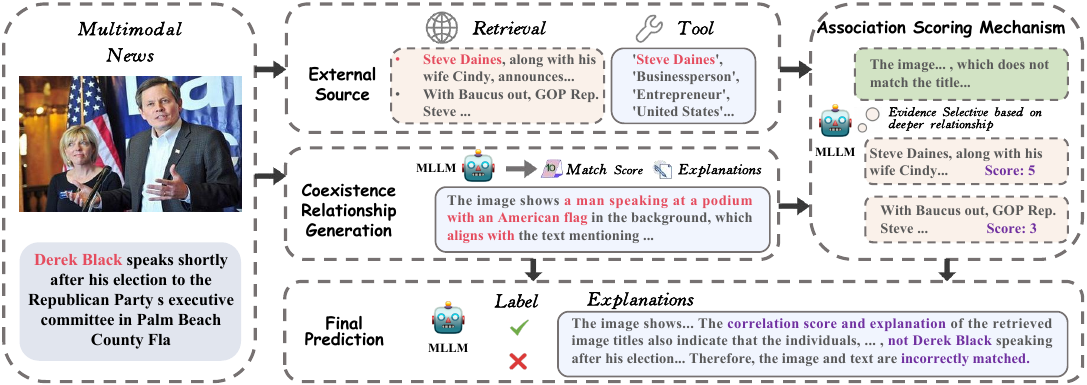}
    \caption{Overall architecture of the proposed CMIE, which effectively models the relationship between external evidence, generating predictive labels and interpretable explanations.}
    \label{framework}
\end{center}
\end{figure*}

\section{Methodology}
Based on the conclusions drawn from the the above exploratory analysis, we introduce CMIE: \textbf{C}ombining \textbf{M}LLM \textbf{I}nsights with \textbf{E}xternal Evidence for Explainable Out-of-Context Misinformation Detection.
As illustrated in Figure  \ref{framework}, the framework takes multimodal news as input and produces final predictive label and explanation.
In our approach, when relevant evidence is unavailable, we rely on the internal knowledge of the MLLM for judgement. When evidence can be retrieved, we focus on rationally integrating the evidence into MLLM's judgement process, which involves two key components: coexistence relationship generation strategy and association scoring mechanism.  The first component identifies the deeper connection of image-text pair, while the second evaluates the significance of the evidence in relation to the connection.

\subsection{Evidence Collection}
When processing a image-text pair, we first need to retrieve relevant evidence. In CMIE, we select the titles associated with the image and the entities extracted from the image as evidence. Our evidence collection pipeline follows the method described in CCN \cite{CCN}. For image titles, we use a web crawler to retrieve the pages where the image appears and save the titles. For entities, we use the Google Vision API \footnote{https://cloud.google.com/vision/docs/detecting-web} to perform extraction. The API returns a list of entities associated with the image, which may describe the content of the image and further describe the context of where these images appear. However, in rare cases where such evidence is not available, we rely on the MLLM’s internal knowledge to directly assess the consistency of the image-text pair and produce a human-readable explanation.

\subsection{Coexistence Relationship Generation}
Our analysis reveals that, after incorporating external evidence, MLLM struggle to identify deeper connection of image-text pair. Instead, it focus on whether the image-text pair align with the evidence. In OOC misinformation, the image's role is to enhance the credibility of the text and should have potential relationship with the text. The proper process for OOC misinformation detection should prioritize the image-text relationship, using evidence to support or refute this relationship. Inspired by this, we propose a coexistence relationship generation strategy that identifies the reasons why image and text might coexist (including direct and potential correlation). Specifically, given an image-text pair $(I, T)$, we use MLLM to first identify and extract potential coexistence relationship $R_{co}$ and coexistence score $S_{co}$. The formulas are as follows:

\begin{equation}
    (R_{co}, S_{co}) = \text{MLLM}(I, T),
\end{equation}
here, $R_{co}$ represents ``why the image and text can appear together", and $S_{co}$ indicates the extent to which image and text can appear together. We follow the paradigm of \textit{(M)LLM-as-a-judge}\cite{gu2024survey, chen2024mllm} to prompt the model to generate corresponding scores without providing explicit definitions.

\subsection{Association Scoring Mechanism}

For retrieved evidences, they can be strongly related, weakly related, or even unrelated. Considering human thought processes, we should prioritize extracting useful information from strongly related evidence to assist the judgment process. Weaker related evidence can be used as supplementary information, helping to further validate the alignment between the image and text.

Therefore, we propose a association scoring mechanism to effectively evaluate the relevance of evidence for the final judgment. Given the coexistence relationship \( R_{co} \) and the set of evidence \( E_t = \{E_{t1}, E_{t2}, \ldots, E_{tn}\} \), we calculate the relevance score $S_i$ and scoring explanation $exp_i$ for each piece of evidence according to the coexistence relationship:
\begin{equation}
    (s_i, exp_i) = \text{MLLM}(R_{co}, E_{ti}),
\end{equation}
By considering both the relevance score and scoring explanation, our approach effectively models evidence and image-text pair. Strongly related evidence significantly influence the final judgment, while weaker related evidence play a supplementary role. This approach ensures that various types of evidence are utilized effectively in evaluating the consistency between the image and text, thereby improving the accuracy and reliability of the judgment.

\subsection{Final Judgement Prediction}

To determine the final prediction for an image-text pair, we integrate the coexistence relationship \( R_{co} \), relevance score \( S_i \), evidence set \( E_{t} = \{E_{t1}, E_{t1}, \ldots, E_{tn}\} \), and image-related entities \( E_{e} = \{E_{e1}, E_{e1}, \ldots, E_{em}\} \) into the MLLM. The final prediction label \( Y_p \) and the corresponding explanation \( Y_e \) are generated as follows:

\begin{equation}
    (Y_p, Y_e) = \text{MLLM}(R_{co}, \{S, exp, E_{t}\},\{E_{e}\}, (I, T)).
\end{equation}

Here, \( Y_p \) is the final prediction label, indicating the correctness of the image-text pair match, and \( Y_e \) provides the explanation for this prediction. Incorporating entities \( E_e \) extracted from image help to contextualize the evidence and refine the analysis, enhancing the accuracy and transparency of the final judgment.

\section{Experiments}
\subsection{Experimental Setup}
\subsubsection{Dataset}We evaluate CMIE and all baselines on the largest benchmark NewsCLIPpings \cite{DBLP:conf/emnlp/newsclippings}, as introduced in exploratory Study.
\subsubsection{Baselines}We compared the proposed method with the following state-of-the-art methods. 
\textbf{CLIP} \cite{DBLP:conf/icml/CLIP}  was fine-tuned on image-caption classification task by adding a linear classifier on the top.
\textbf{DT-transformer} \cite{DBLP:conf/mir/DT-transformer} use CLIP to separate extract image and text feature, and adds a transformer layer to enhance the features interaction. 
The above methods rely on the internal knowledge of the model to complete the judgement.
\textbf{CCN} \cite{CCN} uses external evidence as supporting information, performs circular consistency check on the image-text pair as well as on the external evidence.
\textbf{SEN} \cite{DBLP:conf/emnlp/SEN} propose a stance extraction network, modeling the stances of evidence from different modalities and fully leverages the roles of various types of stance.
\textbf{SNIFFER} \cite{sniffer} utilizes explanations generated by GPT-4 to fine-tune a InstructBLIP \cite{DBLP:conf/nips/InstructBLIP}, and direct integrates evidence to generated predictions.
\textbf{LEMMA} \cite{LEMMA} enhances the detection capabilities of MLLM by retrieving more relevant evidence through a query-based generation method.
The judgement of such methods combines internal knowledge with external evidence.
\subsubsection{Implementation Details} For CLIP and DT-Transformer, feature extraction component is ViT/B-32. For SNIFFER, we directly use the results reported in their paper. Our method is build upon the OpenAI API interface \footnote{https://api.openai.com/v1/chat/completions} and temperature is set to 0.1.


\subsection{Main Results}

Table. \ref{general_results} reports the performance comparison between our proposed method and baselines, along with results from our exploratory experiments. Based on these results, we observe the following:
\textbf{1)} Compared to previous methods that not generate human-readable explanations, MLLM relying solely on internal knowledge to assess inconsistencies between image and text outperform CLIP and DT-transformer but lag behind CCN and SEN, which incorporate external evidence. This indicates that, despite MLLM has the potential to detect OOC misinformation via world knowledge, external evidence remains crucial.
\textbf{2)} After incorporating evidence, the performance of MLLM is comparable to CCN and SEN, which also use external evidence, but falls short of SNIFFER. This suggests that while MLLM augmented by evidence but without fine-tuning has ability to detect OOC misinformation, simply combining MLLM with evidence does not achieve optimal performance. This because the MLLM can not fully grasp the core thread or effectively model the relationship between evidence and the misinformation to be verified.
\textbf{3)} By using CMIE to identify core thread and thoroughly consider the importance of evidence, performance improves significantly, surpassing almost all baselines. This demonstrates the importance of effectively utilizing evidence and optimizing the reasoning process. CMIE effectively filters out noise and highlights important evidence related to core thread, thereby enhancing the MLLM's judgment accuracy.

\begin{table}[t]
\renewcommand\arraystretch{1.4}
\setlength{\tabcolsep}{9 pt} 
\small
  \begin{center}
    \caption{Performance comparison of CMIE and representative baselines on the NewsCLIPpings benchmark.}
\begin{tabular}{cccc}
\hline \hline
 \multirow{2}{*}{\textbf{Method}} & \multirow{2}{*}{\textbf{Accuracy}} & \multicolumn{2}{c}{\textbf{Precision}} \\ \cline{3-4}
  & & \textbf{Real} & \textbf{Fake} \\ \hline
  \multirow{1}{*}{CLIP} & 0.65 & 0.65 & 0.64 \\
  \multirow{1}{*}{DT-Transformer} & 0.78 & 0.77 & 0.79 \\
  \multirow{1}{*}{CCN} & 0.84 & 0.84 & 0.85 \\ 
  \multirow{1}{*}{SEN} & 0.87 & 0.86 & 0.88 \\  
  \multirow{1}{*}{SNIFFER} & 0.88 & \textbf{0.91} & 0.87 \\
  \multirow{1}{*}{LEMMA} & 0.81 & 0.74 & 0.92 \\\cline{1-4}
  \multirow{1}{*}{DR} & 0.83 & 0.79 & 0.87 \\
  \multirow{1}{*}{AR} & 0.85 & 0.86 & 0.83 \\
  \multirow{1}{*}{CMIE} & \textbf{0.91} & 0.88 & \textbf{0.93} \\ \hline \hline
\end{tabular}
    \label{general_results}
  \end{center}
\end{table}

\subsection{Ablation Study}
Table. \ref{ablation} presents the precision performance of our method with various component combinations (Image Title, coexistence Relationship Generation(CRG), Association Scoring (AS), and Entities). 

\begin{table}[h]
\renewcommand\arraystretch{1.5}
\setlength{\tabcolsep}{3.5 pt} 
\small
  \begin{center}
    \caption{Ablation study of CMIE components in terms of accuracy (\%).}
\begin{tabular}{cccc|ccc}
\hline \hline
ImageTitle & CRG & AS & Entity & All & Real & Fake \\ \hline
\XSolidBrush & \XSolidBrush & \XSolidBrush & \XSolidBrush & 0.83 & 0.79 & 0.87 \\
\textcolor{magenta}{\ding{51}} & \XSolidBrush & \XSolidBrush & \XSolidBrush &  0.85 & 0.86 & 0.83 \\
\textcolor{magenta}{\ding{51}} & \textcolor{magenta}{\ding{51}} & \XSolidBrush & \XSolidBrush & 0.87 & 0.83 & 0.91 \\
\textcolor{magenta}{\ding{51}} & \textcolor{magenta}{\ding{51}} & \textcolor{magenta}{\ding{51}} & \XSolidBrush & 0.90 & 0.87 & 0.94 \\
\textcolor{magenta}{\ding{51}} & \textcolor{magenta}{\ding{51}} & \textcolor{magenta}{\ding{51}} & \textcolor{magenta}{\ding{51}} & 0.91 & 0.88 & 0.93 \\
\hline \hline

\end{tabular}
    \label{ablation}
  \end{center}
\end{table}

When using direct reasoning, the accuracy of real samples is lower compared to fake samples. This occurs because the model tends to produce conservative outputs, classifying fake samples without clear errors as real.
After adding the title evidence, the accuracy of real samples decreases because the verification core thread becomes unclear, leading to more real samples being misclassified as fake.
Incorporating CRG significantly improves the accuracy of fake samples but does not enhance the accuracy of real samples due to evidence noise.
With the addition of the AS, which evaluates the importance of evidences based on coexistence relationships, overall accuracy improves. This demonstrates the effectiveness of combining CRG and AS.
Further inclusion of entities extracted from the image results in a slight overall performance improvement, indicating that accurate entity information also contributes to OOC misinformation detection.

In conclusion, the progressive addition of each component enhances performance. The combination of CRG, AS, and entity information significantly improves misinformation detection accuracy, emphasizing the importance of leveraging multiple information sources in a rational manner.

\subsection{Adaptability to Different MLLMs}

\begin{figure}[h]
\begin{center}
    \includegraphics[width=1\linewidth]  {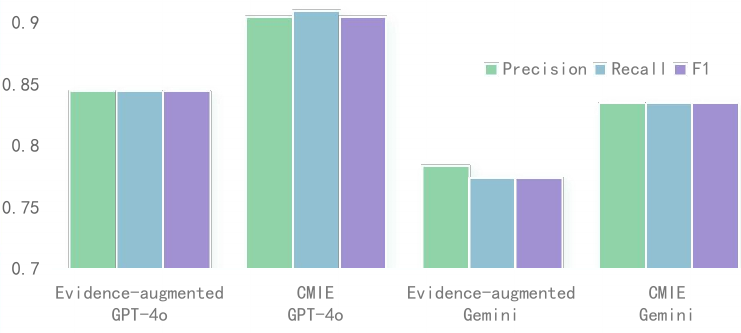}
    \caption{Performance comparison of GPT-4o and Gemini-Pro under evidence-augmented reasoning and CMIE.}
    \label{various_models}
\end{center}
\end{figure}

In addition to GPT-4o, we also evaluated the performance improvements brought by CMIE compared to evidence-augmented reasoning on Gemini-pro. As shown in Figure  \ref{various_models}, CMIE enhances performance for both models. However, due to the inherent limitations in understanding and reasoning capabilities, Gemini-pro's overall performance in the OOC misinformation detection task is lower than GPT-4o.

\subsection{Model Stability Analysis}
Our framework adopts a multi-stage pipeline, raising the potential risk of error propagation, where inaccuracies in earlier stages may compromise downstream performance. To assess the stability of our approach under such risks, we conducted an analysis focused on errors introduced during Stage 1: coexistence relationship generation. We observed two common types of errors at this stage: (i) misinformation instances misclassified as having strong coexistence relationships when they should be weak, and (ii) true information instances misclassified as weak when they should be strong. According to our statistics, 1,278 misinformation samples were affected by such errors, compared to only 36 true information samples, with misinformation cases accounting for 97\% of all error-propagated instances.

Table~\ref{error_prop} presents the performance of our method on the affected misinformation samples. While some degradation in recall and F1 score is observed, the overall performance remains within an acceptable range. These results suggest that, despite upstream errors, the association scoring mechanism in Stage 2 can still effectively leverage relevant evidence, thus contributing to the overall robustness of the framework.

\begin{table}[h]
\renewcommand\arraystretch{1.2}
\setlength{\tabcolsep}{10 pt} 
\small
  \begin{center}
    \caption{Detection performance on misinformation instances impacted by coexistence relationship misclassification in Stage 1. The results indicate that CMIE effectively mitigates upstream errors through the association scoring mechanism in Stage 2.}
\begin{tabular}{ccc}
\hline \hline
Precision & Recall & F1 \\  \hline
0.98 & 0.70 & 0.82  \\ \hline \hline

\end{tabular}
    \label{error_prop}
  \end{center}
\end{table}

\subsection{Robustness to Prompt Variations}
To evaluate the impact of different prompts on model performance, we applied two modifications to the original prompt: task description rewriting and label reversal (see Appendix \ref{app:prompt_sens} for details). This design aims to assess the model’s understanding of the task and evaluate the stability of CMIE. Under these two settings, the model achieved 88\% and 89\% accuracy, respectively, indicating that although there are slight performance differences across prompts, the overall performance remains largely stable.

\subsection{Explanation Quality Assessment}

\label{human_evaluation}
We conducted human assessments to evaluate the quality of the explanations generated by MLLM. We randomly selected 50 image-text pairs and invited ten evaluators to participate. The evaluators rated the explanations generated by MLLM on a scale from 1 to 5, with prior knowledge of the labels (see Appendix \ref{app:human_eval} for details). Higher scores were given if the evaluators considered the generated explanations to be more accurate and comprehensive; otherwise, lower scores were assigned.

\begin{figure}[h]
\begin{center}
    \includegraphics[width=0.8\linewidth]  {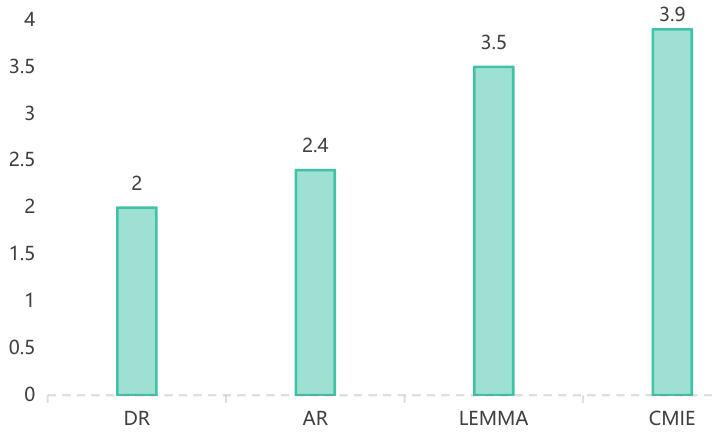}
    \caption{Human evaluation scores for OOC detection explanation quality (scale: 1–5).}
    \label{human_eval}
\end{center}
\end{figure}

\begin{figure*}[t!]
\begin{center}
    \includegraphics[width=1\linewidth]  {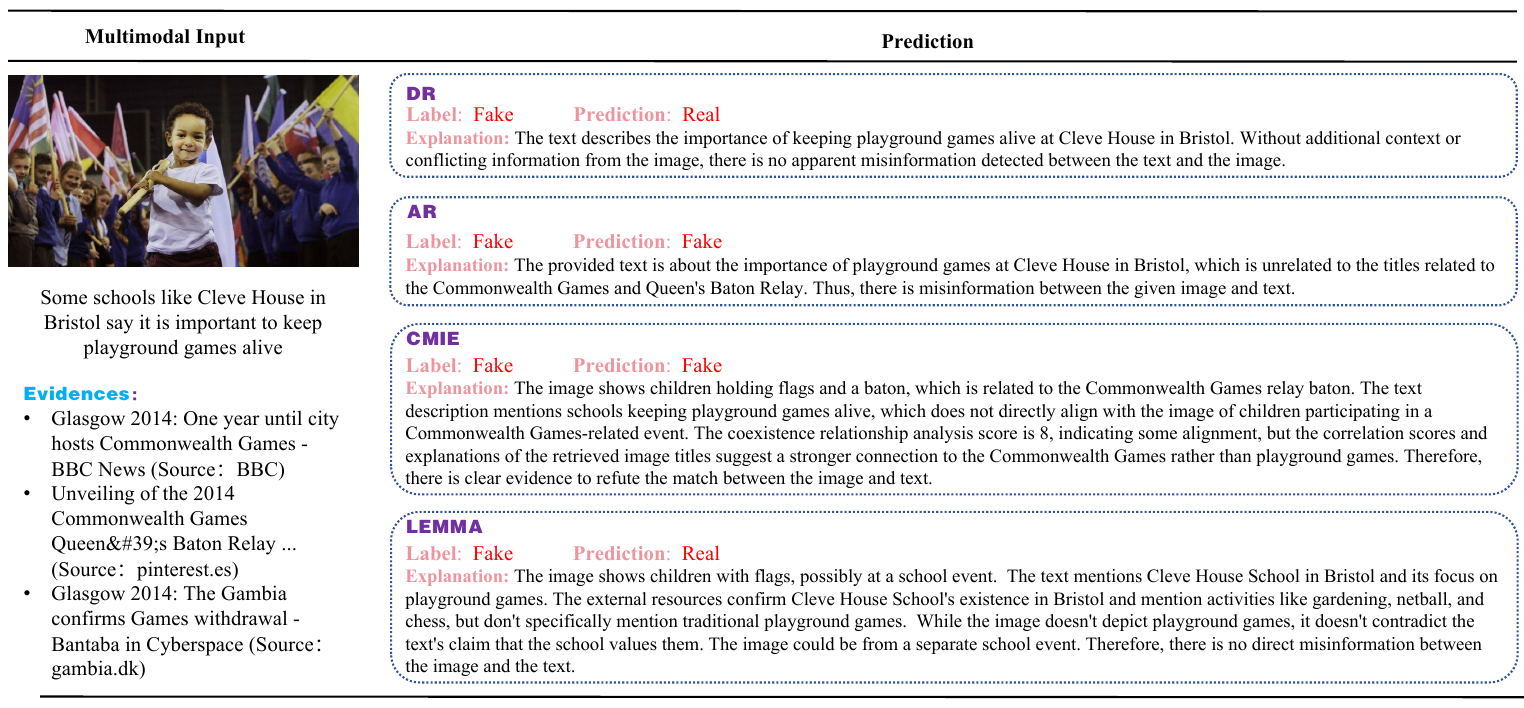}
    \caption{Examples of predictions and explanations generated by the proposed CMIE.}
    \label{case_study}
\end{center}
\end{figure*}

The experimental results in Figure  \ref{human_eval} show that the explanations generated by LEMMA and CMIE outperform those of DR and AR, with CMIE performing better than LEMMA, indicating that CMIE can produce more accurate and comprehensive explanations.

\subsection{Case Study}
Figure  \ref{case_study} presents examples of explanations generated by different methods. As shown in this figure, CMIE and LEMMA can generate more comprehensive explanations, but LEMMA is affected by evidence and fails to identify deep connections, whereas CMIE is able to uncover deeper information, resulting in more accurate outcomes.

\section{Related Work}
\subsection{Multimodal Misinformation Detection}
In the field of OOC misinformation detection, early studies \cite{DBLP:conf/icml/CLIP, DBLP:conf/mir/DT-transformer} utilise pre-trained models to perform out-of-context image-text pair detection. i.e. CLIP. These methods detect misinformation using the internal knowledge of pre-trained models. 
Other works \cite{CCN, DBLP:conf/mir/Muller-BudackTD20} aid the detection of misinformation by introducing external evidence, but such approaches do not take into account the relevance of the evidence to the task. 
SEN\cite{DBLP:conf/emnlp/SEN} indicate that evidence stance represents a bias against results and propose a stance extraction network, modeling the stances of evidence from different modalities and fully leverages the roles of various types of stance.
However, unlike our approach, none of these methods can generate human-readable explanations, which somewhat diminishes the confidence of humans to rely on such methods for decision-making.

\subsection{Retrieval Augmented Generation for OOC Misinformation Detection}
Despite the significant potential demonstrated by LLMs and MLLM across various tasks \cite{DBLP:journals/ftcgv/MLLM_survey1, DBLP:journals/corr/MLLM_survey2}, they still struggle to accomplish complex tasks \cite{DBLP:journals/corr/not_good_fact_checker, DBLP:journals/corr/bad_2} due to their inability to timely update knowledge and their limited understanding of domain-specific knowledge. Research has shown that leveraging Retrieval-Augmented Generation (RAG) can enhance the performance of LLMs \cite{DBLP:journals/corr/RAG_LLM, DBLP:conf/emnlp/WangS23a, DBLP:journals/corr/RAG_check_survey}. By combining retrieved external information such as wikipedia and internet, RAG is able to generate more accurate outputs.
For OOC misinformation detection, existing studies \cite{sniffer, LEMMA, MUSE} also employ Retrieval-Augmented Generation (RAG) techniques to assist in detecting misinformation by retrieving relevant external evidence. Among these studies, SNIFFER \cite{sniffer} requires fine-tuning based on InstructBLIP \cite{DBLP:conf/nips/InstructBLIP} to generate explanations and relies on ChatGPT \cite{DBLP:journals/corr/GPT-4} to generate explanation labels. This method demands significant computational resources and a large amount of labeled data, making it both time- and resource-intensive. In contrast, LEMMA \cite{LEMMA} and MUSE \cite{MUSE} do not require model fine-tuning. These methods also consider the relevance of external evidence to the task. However, they primarily focus on superficial lexical matching between the retrieved results and the query, lacking deeper alignment between the evidence and the image-text pair. When faced with complex OOC misinformation, they may introduce irrelevant or misleading information.

\section{Conclusion}
This paper explores and analyzes the performance of multimodal large language models in out-of-context misinformation detection, both using the internal knowledge and in combination with external evidence. We found that while MLLM exhibit significant potential for OOC misinformation detection, it tend to overlook key validation aspects and susceptible to noise in external evidence. Based on these findings, we proposed the coexistence relationship generation strategy and the association scoring mechanism, which are designed to help MLLM identify deep connections of image-text pair and effectively utilize relevant evidence. Experimental results demonstrate that our method outperforms baselines in most cases. Our research provides a foundation for using MLLM combined with external evidence for OOC misinformation detection, aiding in the development of more effective detection methods.



\section*{Limitations}
The known limitations and possible solutions (plans) of this work are as follows:
\begin{itemize}
\item MLLM suffer from the hallucination problem, which may affect the detection results. To address this issue, our future work plans to further analyze the generated explanations and apply methods to reduce hallucinations.
\item Like other MLLM-based methods, some of the explanations generated by CMIE differ from human evaluation standards, even though it mimics the human recognition process. We will introduce intelligent agents in future work to further improve this consistency.
\end{itemize}

\section*{Ethical Considerations}
This work aims to explore the difficulty of detecting multimodal misinformation using MLLM. Therefore, our work can have a positive impact by mitigating the harm caused by the spread of misinformation in society.

The proposed CMIE relies on the capabilities of MLLM. Therefore, CMIE may produce explanations that deviate from human judgment in certain cases. This is an inherent drawback of MLLM, rather than a limitation of our method.

\section*{Acknowledgement}
This work is supported by the Yunnan Province expert workstations (Grant No: 202305AF150078), National Natural Science Foundation of China (Grant No: 62162067), Yunnan Fundamental Research Project (Grant No: 202401AT070474; 202501AU070059), Yunnan Province Special Project (Grant No:202403AP140021), and Yunnan Provincial Department of Education Science Research Project(Grant No: 2025J0006), and Scientific Research and Innovation Project of Postgraduate Students in the Academic Degree of Yunnan University (KC-4248590).

\bibliographystyle{acl_natbib}

\clearpage
\appendix

\section{Appendix}
\label{sec:appendix}

\subsection{Prompt Sensitive Evaluation}
\label{app:prompt_sens}

To evaluate the impact of different prompts on model performance, we applied two types of modifications to the prompt used for the final prediction: task description rewriting and label reversal.
For task description rewriting, we modified the task instruction by rephrasing it.

Below is the original task description:

\begin{tcolorbox}
[colback=black!2!white,colframe=white!50!black,boxrule=0.5mm]
You will receive an image and a text description, and your task is to determine if the image matches the text description. In addition, you will receive additional auxiliary information, including titles retrieved from images and matching analysis lists with images, entities extracted from images, and a simple coexistence relationship analysis. Coexistence relationship analysis shows a simple reason why images and text are matched together. Please combine other information and a given confidence score to comprehensively determine whether the given image and text are incorrectly matched.
\end{tcolorbox}

And below is the rewritten version:

\begin{tcolorbox}
You are provided with an image and a corresponding text description. Your task is to determine whether the image and the text are appropriately matched. To assist in this judgment, you will also receive auxiliary information, including: image-retrieved titles, image-based matching analysis lists, entities extracted from the image, and a brief analysis of their coexistence relationship—which offers a basic rationale for the image-text pairing. Please evaluate all available information, along with a given confidence score, to make a comprehensive judgment on whether the image and text are mismatched.
\end{tcolorbox}

For label reversal, we modified the output format. In the original prompt, the model was instructed to output "Yes" when misinformation was detected. With label reversal, however, the model is prompted to output "No" under the same condition. This modification helps evaluate the model's task comprehension and stability.

Below is the original output format:

\begin{tcolorbox}
Generate a JSON object with two properties: 'label', 'explanation'. 
The return value of 'label' property should be selected from ["Yes", "No"].
\colorlightblue{Yes} indicates \colorlightblue {there is misinformation} between the given image and text.
\colorlightblue{No} indicates that \colorlightblue {there is no misinformation} between the given image and text.
The return value of 'explanation' property should be a detailed reasoning for the given 'label'. Please provide detailed steps for judging based on scores and other evidence.
\end{tcolorbox}

And below is the label reversal version:

\begin{tcolorbox}
Generate a JSON object with two properties: 'label', 'explanation'. 
The return value of 'label' property should be selected from ["Yes", "No"].
\colorlightblue{Yes} indicates \colorlightblue {there is no misinformation} between the given image and text.
\colorlightblue{No} indicates that \colorlightblue {there is misinformation} between the given image and text.
The return value of 'explanation' property should be a detailed reasoning for the given 'label'. Please provide detailed steps for judging based on scores and other evidence.
\end{tcolorbox}

\subsection{Human Evaluation Details}
\label{app:human_eval}
We conducted human evaluation in Section \ref{human_evaluation}. Below are the instructions provided to the participants.

\begin{tcolorbox}
You will be presented with a multimodal news item consisting of an image and its corresponding caption. Additionally, you will see the news label (true or false) along with the model-generated justification for this label. Please evaluate the justification based on its reasonableness and comprehensiveness, scoring it on a scale from 1 to 5, where a higher score indicates that you find the model’s justification more thorough and convincing.
\end{tcolorbox}

\subsection{Prompts}
The prompts used in the CMIE reported in the following tables respectively. The \colorlightgreen{\{$\cdot$\}} in prompt represent the input content.

\begin{figure*}[h]
\begin{prompt}{Direct Reasoning}
\small  
\texttt{[Guidance]} \\
\texttt{You are given an \colorlightblue{image} and a piece of \colorlightblue{text}. Your task is to \colorlightblue{predict whether there is misinformation} between the given image and text.} \\
\\
\texttt{[Input]} \\
\texttt{The given text: \colorlightgreen{\{text\}}} \\
\\
\texttt{[Output Format]}\\
\texttt{Generate a JSON object with two properties: 'label', 'explanation'. 
The return value of 'label' property should be selected from ["Yes", "No"].
Yes indicates there is misinformation between the given image and text.
No indicates that there is no misinformation between the given image and text.
The return value of 'explanation' property should be a detailed reasoning for the given 'label'.
Note that your response will be passed to the python interpreter, SO NO OTHER WORDS! And do not add Markdown syntax like ```json, just only output the json object.}  \\
\\
Your Response: 
\end{prompt}
\caption{ Prompt of direct reasoning using MLLM.}
\label{fig:direct_reasoning}
\end{figure*}
\begin{figure*}[t]
\begin{prompt}{Evidence Augmented Reasoning}
\small  
\texttt{[Guidance]} \\
\texttt{You are given an \colorlightblue{image}, a piece of \colorlightblue{text}, and a list of \colorlightblue{titles} related to the image content. Your task is to \colorlightblue{predict whether there is misinformation} between the given image and text.} \\
\\
\texttt{[Input]} \\
\texttt{The given text: \colorlightgreen{\{text\}}} \\
\\
\texttt{[Output Format]}\\
\texttt{Generate a JSON object with two properties: 'label', 'explanation'. 
The return value of 'label' property should be selected from ["Yes", "No"].
Yes indicates there is misinformation between the given image and text.
No indicates that there is no misinformation between the given image and text.
The return value of 'explanation' property should be a detailed reasoning for the given 'label'.
Note that your response will be passed to the python interpreter, SO NO OTHER WORDS! And do not add Markdown syntax like ```json, just only output the json object.}  \\
\\
Your Response: 
\end{prompt}
\caption{ Prompt of evidence enhanced reasoning using MLLM.}
\label{fig:augmented_reasoning}
\end{figure*}
\begin{figure*}
\begin{prompt}{Coexistence Relationship Generation}
\small  
\texttt{[Guidance]} \\
\texttt{You are given an \colorlightblue{image}, a piece of \colorlightblue{text}. Your task is to \colorlightblue{provide the reason and score for the coexistence of the picture and text.}} \\
\\
\texttt{[Input]} \\
\texttt{The given text: \colorlightgreen{\{text\}}} \\
\\
\texttt{[Output Format]}\\
\texttt{Generate a JSON object with two properties: 'score','explanation'. 
The return value of 'score' property should be the coexistence degree of image and text , the value of should be selected from the range of [0,10].
The return value of 'explanation' property should be a detailed reasoning.
Note that your response will be passed to the python interpreter, SO NO OTHER WORDS! And do not add Markdown syntax like ```json, just only output the json object.}  \\
\\
Your Response: \\

\end{prompt}
\caption{ Prompt of generating coexistence relationship.}
\label{fig:CRG}
\end{figure*}
\begin{figure*}
\begin{prompt}{Association Scoring Mechanism}
\small  
\texttt{[Guidance]} \\
\texttt{You will receive an \colorlightblue{image}, a \colorlightblue{title} related to the image content, a \colorlightblue{text} description as well as a \colorlightblue{coexistence relationship} about the image and text description. Your task is to \colorlightblue{score the relevance level of each title} based on images,text description and coexistence relationship. Please first decompose the coexistence relationship into several topics to be validated, and rate the title based on these topics combined with images. Please think step by step.   
} \\
\\
\texttt{[Input]} \\
\texttt{The given text: \colorlightgreen{\{text\}}} \\
\texttt{The list of titles related to the image content: \colorlightgreen{\{text\}}} \\
\texttt{The coexistence relationship: \colorlightgreen{\{text\}}} \\
\\
\texttt{[Output Format]}\\
\texttt{Generate a JSON object with three properties:'index', 'score', 'explanation','original title'. 
The return value of 'index' property should be the index of title.
The return value of 'score' property should be the degree of correlation between the title and the image, the value of should be selected from the range of [0,10].
The return value of 'explanation' property should be a detailed reasoning for the given 'score'.
The return value of 'original title' property should be the original text of title.}  \\
\\
Your Response: \\
\end{prompt}
\caption{ Prompt of scoring the image related evidences.}
\label{fig:AS}
\end{figure*}

\begin{figure*}
\begin{prompt}{Final Prediction}
\small  
\texttt{[Guidance]} \\
\texttt{You will receive an \colorlightblue{image} and a \colorlightblue{text} description, and your task is to determine if the image matches the text description. In addition, you will receive additional auxiliary information, including \colorlightblue{titles} retrieved from images and \colorlightblue{matching analysis lists} with images, \colorlightblue{entities} extracted from images, and a simple \colorlightblue{coexistence relationship} analysis. Coexistence relationship analysis shows a simple reason why images and text are matched together. Please combine other information and a given confidence score to comprehensively \colorlightblue{determine whether the given image and text are incorrectly matched}. 
} \\
\\
\texttt{[Input]} \\
\texttt{The given text: \colorlightgreen{\{text\}}} \\
\texttt{The list of titles related to the image content: \colorlightgreen{\{text\}}} \\
\texttt{The coexistence relationship analysis of image and text: \colorlightgreen{\{text\}}} \\
\texttt{The correlation score and explanation of retrieved image titles and images: \colorlightgreen{\{text\}}} \\
\texttt{The the entities extracted from image: \colorlightgreen{\{text\}}} \\
\\
\texttt{[Output Format]}\\
\texttt{Generate a JSON object with two properties: 'label', 'explanation'. 
The return value of 'label' property should be selected from ["Yes", "No"].
Yes indicates there is misinformation between the given image and text.
No indicates that there is no misinformation between the given image and text.
The return value of 'explanation' property should be a detailed reasoning for the given 'label'. Please provide detailed steps for judging based on scores and other evidence.
Note that your response will be passed to the python interpreter, SO NO OTHER WORDS! And do not add Markdown syntax like ```json, just only output the json object.
}  \\
\\
Your Response: \\
\end{prompt}
\caption{ Prompt of generating final decision results.}
\label{fig:final_pred}
\end{figure*}

\end{document}